# Channel Reordering with Time-shifted Streams to Improve Channel Change Latency in IPTV Networks

Aytac Azgin, *Member*, IEEE, and Yucel Altunbasak, *Member*, IEEE

**Abstract** — *In IPTV networks, channel change latency is considered as a major obstacle in achieving broadcast-level quality video delivery. Because of the bandwidth limitations observed at the client side, users typically have immediate access to a limited number of channels. As a result, channel change requests oftentimes need to go through the network, thereby leading to significant delays. In this paper, we address this problem by proposing a resource-efficient time-shifted channel reordering mechanism to minimize the channel change latency. The proposed framework exploits the differing key-frame delivery times for the adjacent sessions to dynamically arrange the switching order during the surfing periods. The simulation results show that, with the proposed framework, more than 50% improvement can be achieved in channel change latency without introducing any overhead in the network.*

## I. INTRODUCTION

Internet Protocol Television (IPTV) is the general term used to describe services, which deliver reliable, secure, and quality-of-service (QoS)-enabled triple-play content over the IP infrastructure [1]. With the increased penetration rate for the broadband access technologies, IPTV has become a serious alternative to the traditional broadcast systems, thereby garnering significant interest from industrial and research communities. As the IPTV content is typically delivered over non-reliable transmission channels, we need explicit framework designs, which are supported with reliable transport techniques, to ensure that the service quality requirements for the IPTV service (i.e., packet loss rate, latency, and jitter) are met at the end users. Our paper focuses on the latency aspect of the service quality assurance, and develops a novel channel change framework to improve the latency performance in IPTV networks.

Channel change latency has a critical importance in evaluating the end-to-end performance of an IPTV network. Therefore, reducing the channel change latency is a major design goal for the IPTV service providers to deliver broadcast quality content to the IPTV clients [2]. To achieve this objective, various approaches have been proposed (for further details, see [3]). We can categorize these approaches into three main groups: (*i*) content-based solutions (*e.g.*, [4, 5]), (*ii*) network-assisted solutions (*e.g.*, [6, 7]), and (*iii*) client-based solutions (*e.g.*, [8, 9]). In general, the majority of the studies performed to improve the channel change latency require modifications at the network level to some extent, either at the content delivery server (that is typically stationed at the Video Hub Office), or near the access network (*e.g.*, through the installation of dedicated servers at the Video Switching Office). In this paper, we propose a client-based solution that requires minimal changes at the content delivery server.

Client-based channel change techniques typically focus on the user preferences to improve the channel switching latency. Specifically, these approaches typically acquire information on channel access frequencies and/or the distributions associated with the conditional access probabilities for the channel switching events by analyzing the users' channel change behaviors. In doing so, a prioritized channel listing can be determined for the surfing periods [1] to deliver concurrent channel change streams to the IPTV clients [10]. Or, we can reorder the channel listing at the client side to minimize the number of switches triggered during a surfing period [11]. If, for instance, the clients switch through the channels using mostly the up/down buttons in their remote control, reordering the channel listings may allow the clients to reach their targeted sessions sooner. If we also combine the channel reordering framework with the concurrent delivery framework, we can significantly improve the channel change latency. However, concurrent delivery techniques are also shown to introduce significant overhead at both the client side and the access network to achieve the minimum required latency objectives at the client side (see [12] for a detailed technical analysis).

The research proposed in this paper utilizes a different approach to integrate the channel reordering concept into an IPTV channel change framework. We specifically focus on the delivery of time-shifted group of picture (GOP) sequences for the adjacent channels. In the literature, time-shifted delivery is a concept that is mainly used for the delivery of additional support streams (*e.g.*, [7]). By applying the same principle on the source streams, we propose to design a highly efficient channel change framework that can achieve the latency objectives without introducing significant overhead at the client side or the access network.

The rest of the paper is organized as follows. In Section II we present our system model. We analyze the performance of the proposed channel change framework in Section III. Section

---

[1] A *surfing period* represents the duration in which the client triggers sequential switching events until it settles on a specific session. If the client watches session $s_i$ and decides to switch to session $s_j$, any session that is accessed by the client in-between $s_i$ and $s_j$-along the selected direction of switch-falls within that period.

IV concludes our paper.

## II. SYSTEM MODEL

We consider an IPTV network with *N* active sessions. The sessions are ordered using a specific circular placement technique (see [11] for details), which distributes the sessions evenly based on their access frequencies. This approach is referred to as *frequency interleaved ordering*, and it can significantly reduce the average number of switches triggered during the sequential switching phase (*i.e.*, when the up/down buttons are pressed in the remote control).

In Figure 1 we illustrate two possible interleaving scenarios that can be used to reduce the number of channel switches during a surfing period. *One-step interleaving* is the approach proposed by Lee *et al.* in [11], which assigns sessions with even-numbered popularity rankings and sessions with odd-numbered popularity rankings in reverse directions (*i.e.*, up or down), starting with the most popular session. The *two-step interleaving* approach, on the other hand, is considered here as an alternative to the one-step approach, which assigns two successive channels along any direction before switching to assigning sessions along the reverse direction.

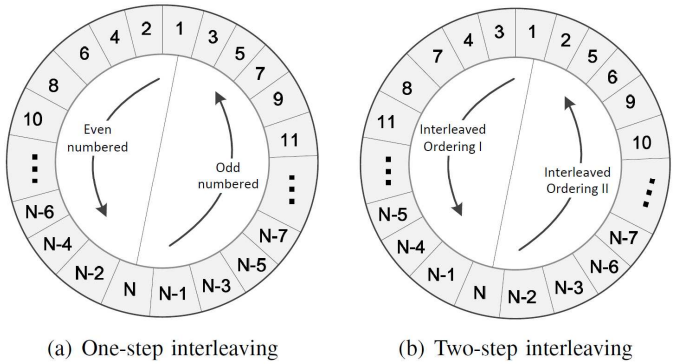

**Fig. 1. Interleaved ordering scenarios (N is assumed to be an even number).**

In Table I we compare the switching performance for the one-step and the two-step interleaving approaches at different *N* values. In our analysis, we assumed the client to switch to channels along the shortest distance towards the targeted session (along either upward or downward direction), instead of randomly choosing the switching direction. We then compared the performances of the one-step and two-step interleaving approaches with regard to the expected number of switches required to reach the targeted session, $E[D_S]$, using the following equation:

$$E[D_S] = \sum_{i=1}^{N} \pi(i) \sum_{\substack{j=1 \\ j \neq i}}^{N} p_{i,j} \times d_{min,ij}, \quad (1)$$

where $\pi(i)$ represents the probability of watching session $s_i$, $p_{i,j}$ represents the probability of switching from session $s_i$ to session $s_j$, and $d_{min,ij}$ represents the minimum switching distance between sessions $s_i$ and $s_j$ (*i.e.*, $d_{min,ij} = min(d_{ij}^{up}, d_{ij}^{dn})$, where $d_{ij}^{up}$ and $d_{ij}^{dn}$ represent the switching distances along the upward and downward directions).

**TABLE I**
**AVERAGE NUMBER OF CHANNEL SWITCHES**

| N | 100 | 200 | 300 | 400 | 500 |
|---|---|---|---|---|---|
| One-step | 15.4497 | 27.9877 | 39.7763 | 51.1233 | 62.1622 |
| Two-step | 15.4541 | 27.9912 | 39.7794 | 51.1262 | 62.1649 |

The results shown in Table I suggest that minor modifications to the interleaved channel ordering have a limited (and oftentimes negligible) impact on the channel switching performance. Based on these observations, we can make the following assumption in regard to the channel reordering: *limited changes to channel switching order during an active surfing phase do not cause noticeable degradations in the time-averaged channel switching performance.* [2]

To take advantage of the assumption considered for the channel reordering process, we propose a novel channel change framework that focuses on the delivery of time-shifted source streams (with respect to the source streams from other sessions) to the clients. To be specific, we introduce a time-shift of *ΔT* between adjacent sessions (*i.e.*, $T_{S,i} = T_{S,i-1} + \Delta T$, where $T_{S,i}$ represents the respective start-of-delivery time for session $s_i$'s key frame packets.).

The idea of using time-shifted streams to support the channel change process in IPTV networks has been previously investigated in various studies. However, these studies have mainly focused on using time-shifted support-streams to improve the latency performance. In this paper, we apply the idea of time-shifted streams directly on the source streams, without using additional support streams, thereby significantly improving the bandwidth efficiency at the access network.

We illustrate the proposed time-shifted delivery process in Figure 2 when $\Delta T = T_{GOP}/4$, where $T_{GOP}$ represents the GOP duration [3]. For the given example, GOP start times for the adjacent sessions are separated by a period of $T_{GOP}/4$ (when compared to a common reference time). Session numbers, here, represent the priority order used during the channel reordering phase (*i.e.*, a session ID of 1 represents the order for the most popular session).

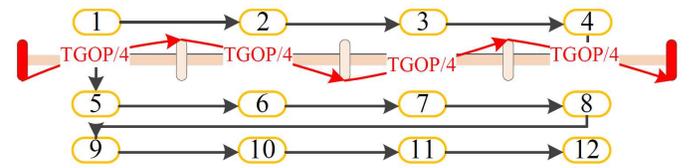

**Fig. 2. Timing for the adjacent sessions.**

The proposed framework utilizes two variables, *ΔT*, the separation interval, and *ΔW*, the maximum waiting period. In

---

[2] Note that, the proposed assumption requires the repetitive loops (such as going back and forth the same session) to be avoided, especially if the modifications are based on a specific reordering algorithm.

[3] Note that, to illustrate the maximum possible gains, we assume the channels to have the same GOP duration. In practice, we only need to apply time-shifts on the key-frame delivery times.

Figure 3 we illustrate a few of the possible step-by-step transitions for the channel change process, which assumes $\Delta T = T_{GOP}/4$ and a waiting period of at most *4* requests. [4] In the given example, the original channel switching order is given by *{1, 2, 3, 4, 5, 6, 7 …}*. Therefore, based on the given parameters, zapping client needs to switch to the *2nd* session by the *4th* request (assuming the client starts switching from the *1st* session). Also note that, because of the selected value for the $\Delta T$ parameter, $T_{S,2}$ equals $T_{S,6}$, which is why session 6 cannot be accessed before session 2 is accessed.

In short, in our framework, the parameter $\Delta W$ ensures that the time-adaptive changes dictated on the initial switching order (due to relative channel switching times) do not propagate beyond a certain limit. In doing so, we can provide fair access to the available sessions with limited delay.

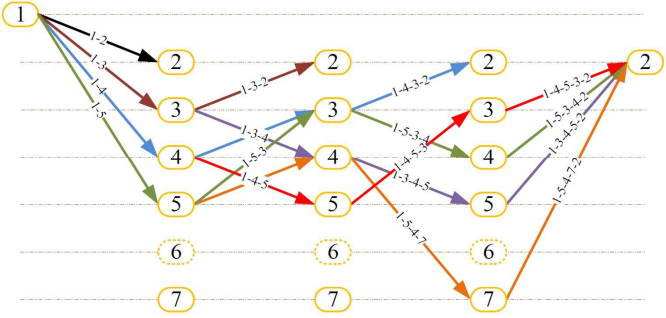

**Fig. 3. Sample paths for the channel change process**

### III. PERFORMANCE ANALYSIS

To analyze the performance of the proposed channel change approach, we created a simulation framework in Matlab to observe for *1x10^6* channel switching events. The results reported here represent the average values.

In our simulations, we have investigated the performance associated with the following metrics: the number of switching events required for reaching the targeted session, and the per-switch latency observed during a surfing period. The parameters that we varied in our simulations are the number of sessions (*N*), maximum wait window size ($\Delta W$), and the separation distance between sessions of equal key-frame start-of-transmission times (*S*, which is given by $T_{GOP}/\Delta T$). We limit the value of *S* to the range of *[3-6]*, and the value of $\Delta W$ to the range of *[2-10]*. The GOP duration is assumed to be equal to *1* second (which suggests a value of $T_{GOP}/2 = 500ms$ for the mean channel change latency, when no additional support stream is used). Reported results are based on the comparisons between the proposed dynamic reordering framework and the frequency interleaved reordering approach.

To represent the channel popularity dynamics in the network, we use the Zipf distribution [13]:

$$\pi(i) = \frac{1/i^s}{\sum_{1 \le n \le N} 1/n^s}, \quad (2)$$

where *π(i)* represents the probability of watching the *ith* most popular session (as defined in Section II) and *s* represents the shape parameter for the given Zipf distribution. In our simulations, we assumed a value of *1* for the shape parameter, i.e., *s=1*.

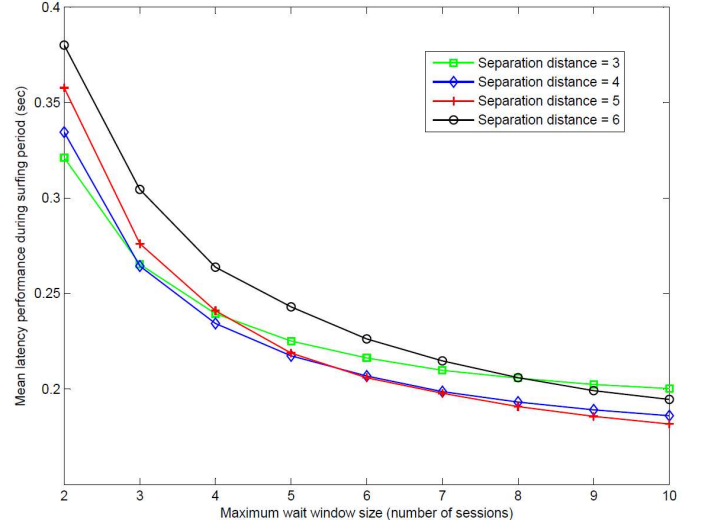

**Fig. 4. Impact of window size on the latency performance**

We start by illustrating the impact of varying the separation distance and $\Delta W$ metrics on the perceived channel change latency. [5] The results are shown in Figure 4.

We observe in Figure 4 that the proposed reordering mechanism improves the latency performance by *30%* when $\Delta W=2$, and by *60%* when $\Delta W=10$. Since increasing the $\Delta W$ value suggests a higher switching count during a given surfing period, for which the results are shown in Figure 5, keeping the value of $\Delta W$ within the *[S-1, S+1]* range seems to provide a good tradeoff between the latency and the switching performances. Note that, even when we use a large value for the $\Delta W$ parameter, the percentile change in the switching performance still stays within an acceptable range. For instance, except for the case when *S=6*, we observe at most a *5%* increase in the mean switching count. Furthermore, when the $\Delta W$ parameter lies within the given range of *[S-1, S+1]*, we can keep the switching overhead to less than *3%*.

Another way to look at the improvements observed for the latency performance is to examine the cumulative distribution function for the perceived latency values corresponding to each channel switching event. We illustrate the results in Figure 6, where we report the results for three scenarios: (*S=3, $\Delta W=2$*), (*S=4, $\Delta W=3$*), and (*S=6, $\Delta W=5$*). We observe that, when *S=3*, around *45%* of the switching events record a latency of *≤250ms*, whereas for the *S=4* and *S=5* cases that ratio drops to *25%* and *10%*, respectively. As we illustrated in Figure 4, by increasing the maximum wait window size, we can further improve those ratios, causing the majority of the requests to observe latency values that are less than the critical threshold of *250ms*.

---

[4] Specifically, the client needs to switch to the channel, which it was supposed to access after the first switch, by the fourth request the latest.

[5] Note that, with latency, we essentially refer to the time delay until the client starts receiving the key-frame packets from the source multicast.

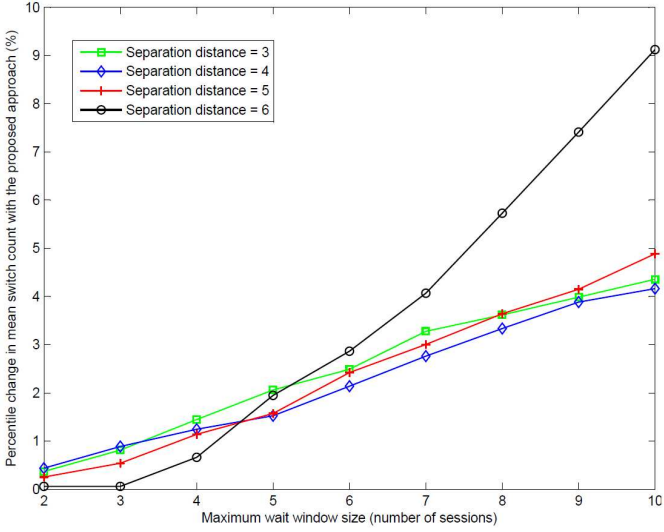

**Fig. 5. Impact of window size on the average number of switches.**

Note that the results presented earlier assume synchronized GOP durations and a common channel order shared by all the users. However, for various reasons, we may not achieve the fully synchronized GOP timings in practice. For instance, clients may use a channel order that reflects their preferences and might be much more different than the initially considered channel order. Or, the sessions may use different GOP durations, which may cause the timings between sessions to change in time.

Therefore, to show the impact of imperfect timings and differing channel orders, we next study the relationship between (pseudo-)randomized channel reordering at the client side and the user perceived latency. For that purpose, we first need to create a channel order that reflects the above suggestions. To do that, we use the following approach, which creates a channel order that partially depends on the network-wide channel access characteristics.

Let us first start with a few definitions: assume that $\pi_N$ represents the probability distribution for the channel watching probability over the whole network and, accordingly, the original channel placement is represented with the set $\Theta$, where $\Theta(i)$ represents the location for the *ith* most popular session along the circular grid. To create a channel order for a specific client, we use a random vector $p$, which consists of $N$ uniformly distributed values from the *[0,1]* range. Assume that the set $N_U$ represents the channels that are not currently assigned to a specific location (*i.e.*, $N_U(i)$ represents the *ith* most popular session among the sessions included in $N_U$). Therefore, during the initialization phase, $N_U = \{N\}$. Also assume that $\Pi_{Nu}$ represents the cumulative distribution array for sessions in $N_U$:

$$\Pi_{N_U}(i) = \sum_{1 \leq j \leq i} \pi_{N_U}(i). \qquad (3)$$

Then, at the first step, we use $max_{loc}\ p(1) \leq \Pi_{N_U}(loc)$ to assign the first channel along the circular grid, *i.e.*, $\Theta^*(1) = loc$, where $\Theta^*$ represents the new channel order. We next update $N_U$ by removing the previously selected session from $N_U$ (*i.e.*, $N_U = N_U \setminus i$), and update $\Pi_{Nu}$ using the updated $N_U$ set. This procedure is recursively repeated until all the available sessions are assigned to a distinct location along the circular channel grid.

Note that, the selection process for the first *N/2* assignments favors the popular channels, whereas the opposite is observed for the last *N/2* sessions. This process is finalized when we update the selection probabilities for the reallocated sessions by mapping the probabilities from the original ordering to the new one (*i.e.*, $\pi^*(i) = \pi(\Theta^{-1}(i))$, where $i^* = \Theta^*(i)$ and $\Theta^{-1}(\Theta(i)) = i$). [6]

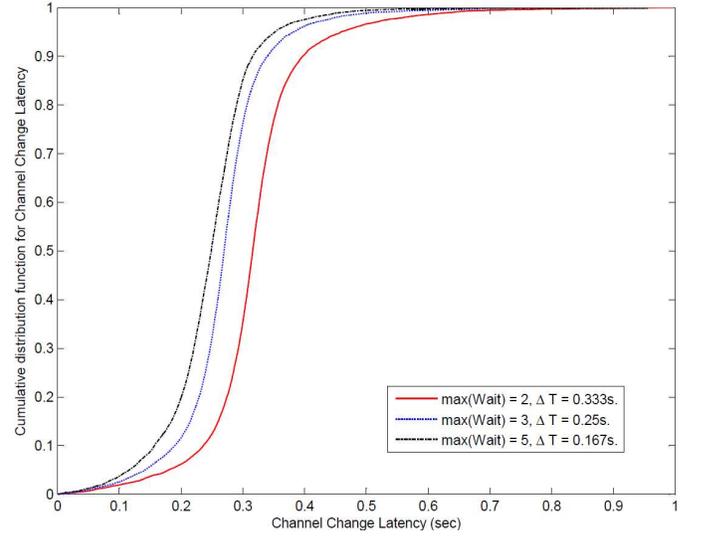

**Fig. 6. Cumulative distribution function for the channel change latency as we vary the duration of separation window.**

We illustrate the latency results that correspond to the pseudo-randomized channel reordering in Figure 7. [7] As expected, we observe an increase in the latency values, since the channel order at the client side does not accurately reflect the initially presumed order (*i.e.*, the time-shifts are originally decided based on the network-wide channel watching routines, which is, here, assumed to be different than the client's watching behavior). Therefore, an IPTV client that uses a channel order that is different than $\Theta$ observes imperfect time-shift assignments for the adjacent sessions in its channel grid.

---

[6] When *N=100*, we observed an Euclidian distance of *0.2627* between the original probability distribution and the pseudo-randomized one. Here, the Euclidian distance measure represents how much more/less likely the channels from the original order are to be watched using the proposed randomized ordering.

[7] Note that we still assume synchronized time-shift values for the given sessions, *e.g.*, if *S=4*, then $|T_{S,i} - T_{S,j}| \in \{0, 0.25s, 0.5s, 0.75s\}$, $\forall \{i,j\} \in \{N\}$.

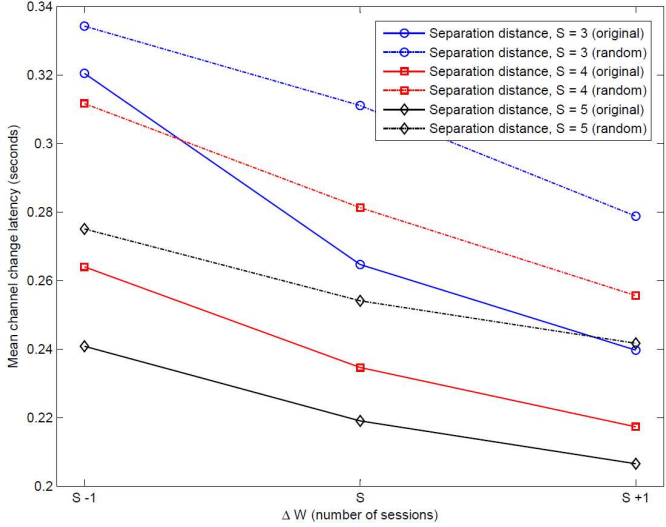

Fig. 7. Impact of randomized session ordering at the client side on channel change latency.

These imperfect assignments lead to around *15*-to-*20%* degradations in the latency performance, suggesting an increase of around *40ms* for the channel change latency perceived by the client, as shown in Table II. However, if we compare our results to that of the frequency interleaved channel ordering approach, the proposed framework is still shown to achieve significant improvements in the perceived latency performance.

We next study the impact of imperfect time-shifts on the channel change performance. For that purpose, we assign each session a randomly selected time-shift value from the [*0s,1s*] range. We illustrate the latency results in Figure 8, for different $S$ and $\Delta W$ values. [8]

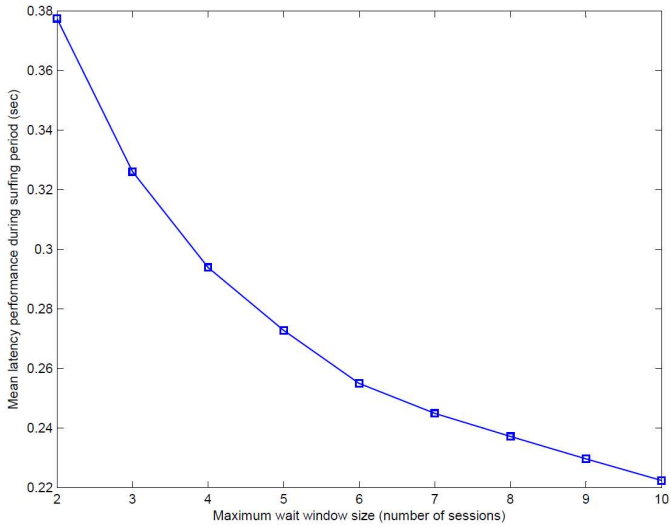

Fig. 8. Impact of randomized time-shifts on channel change latency.

---

[8] In essence, the parameter $S$ does not have any impact on the channel selection process, since the time-shifts are assigned randomly. For this reason, only one set of results are presented for the current scenario.

TABLE II
PERCENTILE CHANGE IN LATENCY WITH RANDOMIZED ORDERING

| $\Delta S$ | 3 | 4 | 5 |
|---|---|---|---|
| $\Delta W = \Delta S - 1$ | 4.31% | 17.53% | 16.31% |
| $\Delta W = \Delta S$ | 18.07% | 19.91% | 17.63% |
| $\Delta W = \Delta S + 1$ | 14.24% | 16.03% | 17.05% |

Once again, we observe noticeable degradation in the latency performance (when compared to the non-randomized time-shifting scenario), which as shown in Figure 9 suggests an increase of around *20%*.

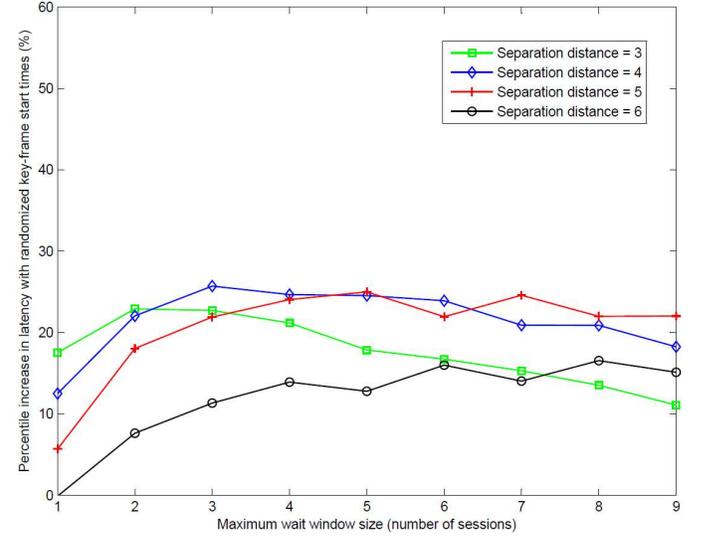

Fig. 9. Percentile change in latency performance due to randomly selected key-frame transmission times.

We also present the results for the accumulative latency observed during a surfing period to better capture the actual performance perceived by the client. We show the results in Figure 10.

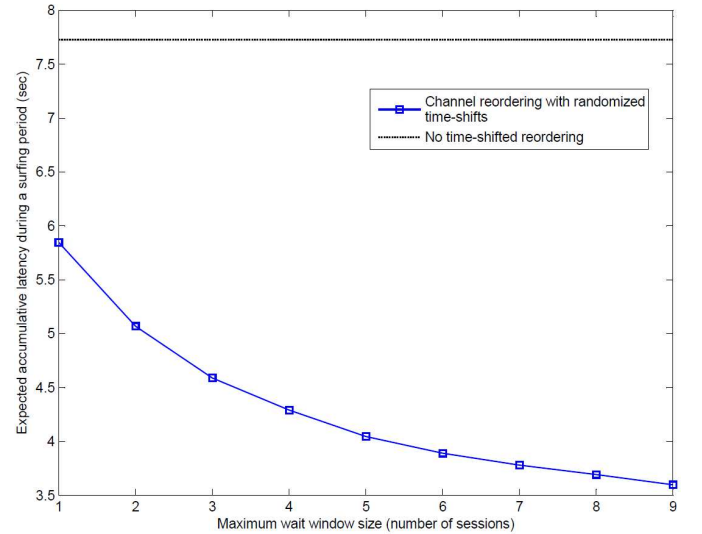

Fig. 10. Timing for the adjacent sessions.

We observe in Figure 10 that by exploiting the time-shift among adjacent sessions to temporarily reorder them during the surfing periods, it is possible to achieve *26*-to-*53%* improvement in the latency performance, which again illustrates the most important aspect of the proposed framework, *i.e.*, achieving significant gains in the latency performance without the need for the delivery of additional support streams, thereby eliminating the excessive overhead incurred at the access network during the surfing periods.

## IV. CONCLUSION

In this paper, we presented a resource-efficient dynamic channel reordering framework that exploits the time-shifts between adjacent sessions to improve the channel change latency. The proposed framework introduces temporary changes to the channel selection order during the channel switching process to minimize per-switch and overall latency values incurred during the surfing periods. We analyzed the performance of the proposed framework using a simulation-based study and the results demonstrated significant improvements in the latency performance without introducing additional overhead in the network.